\documentclass{article}

\PassOptionsToPackage{numbers,sort&compress}{natbib}

\usepackage[final]{neurips_2021_ml4ps}

\usepackage[utf8]{inputenc} 
\usepackage[T1]{fontenc}    
\usepackage{hyperref}       
\usepackage{url}            
\usepackage{booktabs}       
\usepackage{amsfonts}       
\usepackage{nicefrac}       
\usepackage{microtype}      
\usepackage{xcolor}         

\usepackage{graphicx}    
\usepackage{subfig}
\usepackage{amsmath}

\title{Nonlinear pile-up separation with LSTM neural networks for cryogenic particle detectors}

\author{%
  Felix Wagner \\
  Institute of High Energy Physics\\
  Austrian Academy of Sciences\\
  1050 Vienna, Austria \\
  \texttt{felix.wagner@oeaw.ac.at} \\
}

\begin{document}

\maketitle

\begin{abstract}
  In high-background or calibration measurements with cryogenic particle detectors, a significant share of the exposure is lost due to pile-up of recoil events. We propose a method for the separation of pile-up events with an LSTM neural network and evaluate its performance on an exemplary data set. Despite a non-linear detector response function, we can reconstruct the ground truth of a severely distorted energy spectrum reasonably well.
\end{abstract}


\section{Introduction}

In modern experimental particle physics, some of the biggest open questions, such as the search for lightweight dark matter and coherent neutrino nucleus scattering, require the measurement of low energy particle scattering \cite{billard2021direct, doi:10.1063/1.4915590}. Sensitivity to these interactions imposes certain requirements: A low energy threshold for nuclear recoils, low background, a high discrimination power against remaining background and artifacts and a sensible energy calibration down to threshold. Cryogenic detectors, as used by the CRESST, COSINUS and NUCLEUS experiments \cite{PhysRevD.100.102002, Reindl_2020, rothe_nucleus_2020}, meet these requirements, which we shall describe in Sec. \ref{sec:detector}. 

In a measurement, the recoil of a particle within a cryogenic detector causes a pulse-shaped heat signal, which is then triggered and stored within a record window for further analysis. In above-ground and calibration runs, we observe high trigger rates, leading to a large amount of pile up events, i.e. heat signatures of several events within the same record window. If not treated carefully, these events distort the reconstructed recoil energy spectrum significantly. As an alternative to identifying and removing pile up events, we present in our work a method to reconstruct the individual pulse traces and corresponding recoil energies, and by that make use of the measurement's full exposure. The separation is especially challenging if non-linearities in the detector's response function are present, which is usually handled in an individual step of the analysis \cite{stahlberg_2020}. However, with an LSTM neural network \cite{HochSchm97} trained on a simulated data set, we combined these steps into one: We train the model to reconstruct the unsaturated and separated pulse traces from saturated pile-up event. Our results are described in Sec. \ref{sec:results}, a discussion of their implications and future plans are given in Sec. \ref{sec:discussion}. In this work, we used the PyTorch, PyTorch Lightning and Cait Python packages \cite{NEURIPS2019_9015, falcon2019pytorch, wagner_felix_2021_5091416}.

\textbf{Related work.} Machine learning models in rare event search experiments have been mostly applied for the rejection of artifacts and background \cite{Golovatiuk_2020, anais_ml} and the fine-tuning of acceptance regions \cite{Armengaud_2016}. Specifically the rejection of pile up pulses was addressed in Ref. \cite{PhysRevC.104.015501} and \cite{Huang_2021}. There are non-machine learning methods for pile up separation, e.g. in Ref. \cite{PRESTON2021165601}. Contrary to these, we reconstruct the whole pulse trace of the individual events, with significantly faster run time than fitting methods would require, and at the same time treat the nonlinear detector response. We are not aware of any method that reports similar results for the task of nonlinear pile up separation, with comparable inference time.


\section{Detector concept} \label{sec:detector}

\begin{figure}[!t]
\centering
\subfloat[]{\includegraphics[width=0.47\textwidth]{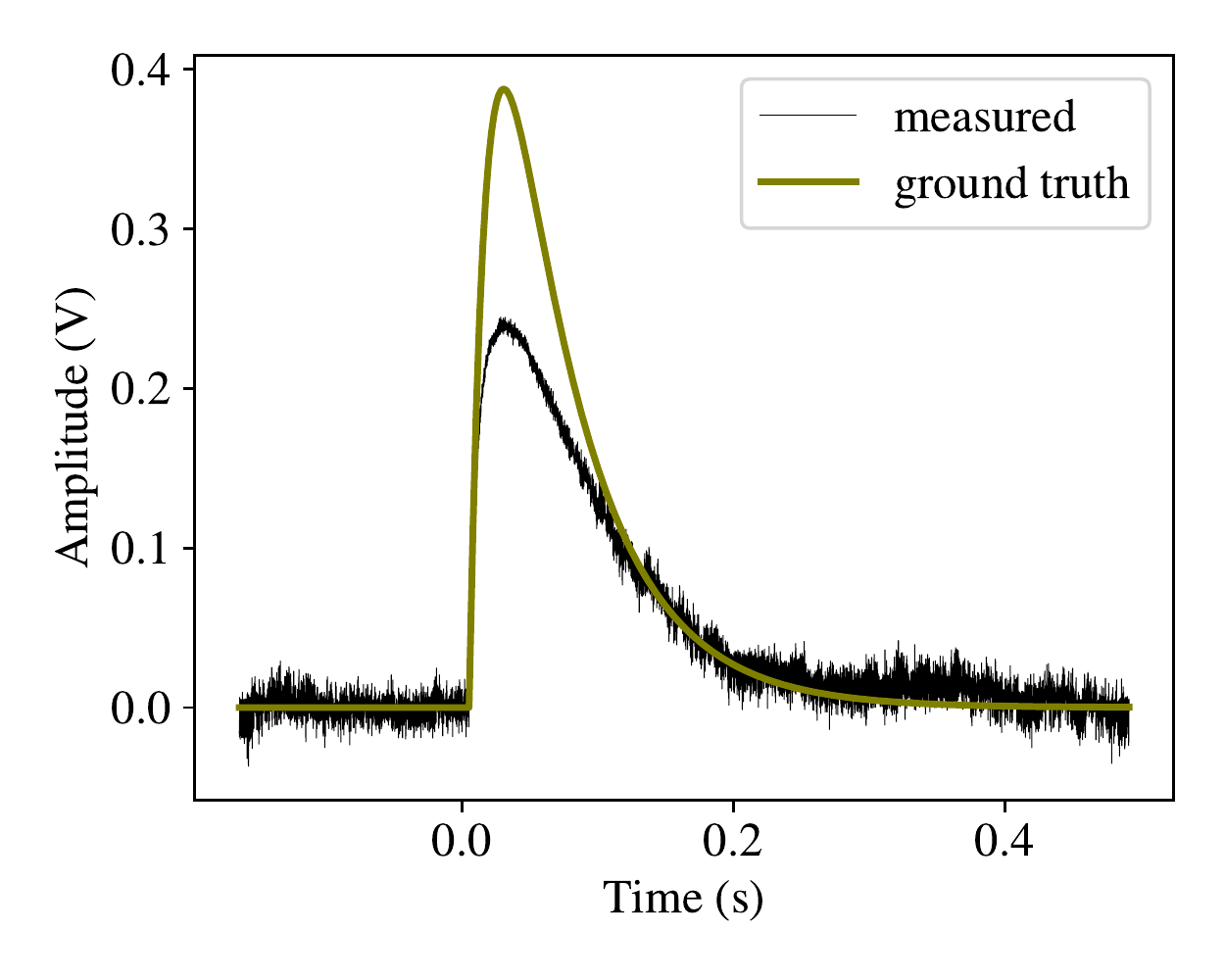}} \quad \quad 
\subfloat[]{\includegraphics[width=0.47\textwidth]{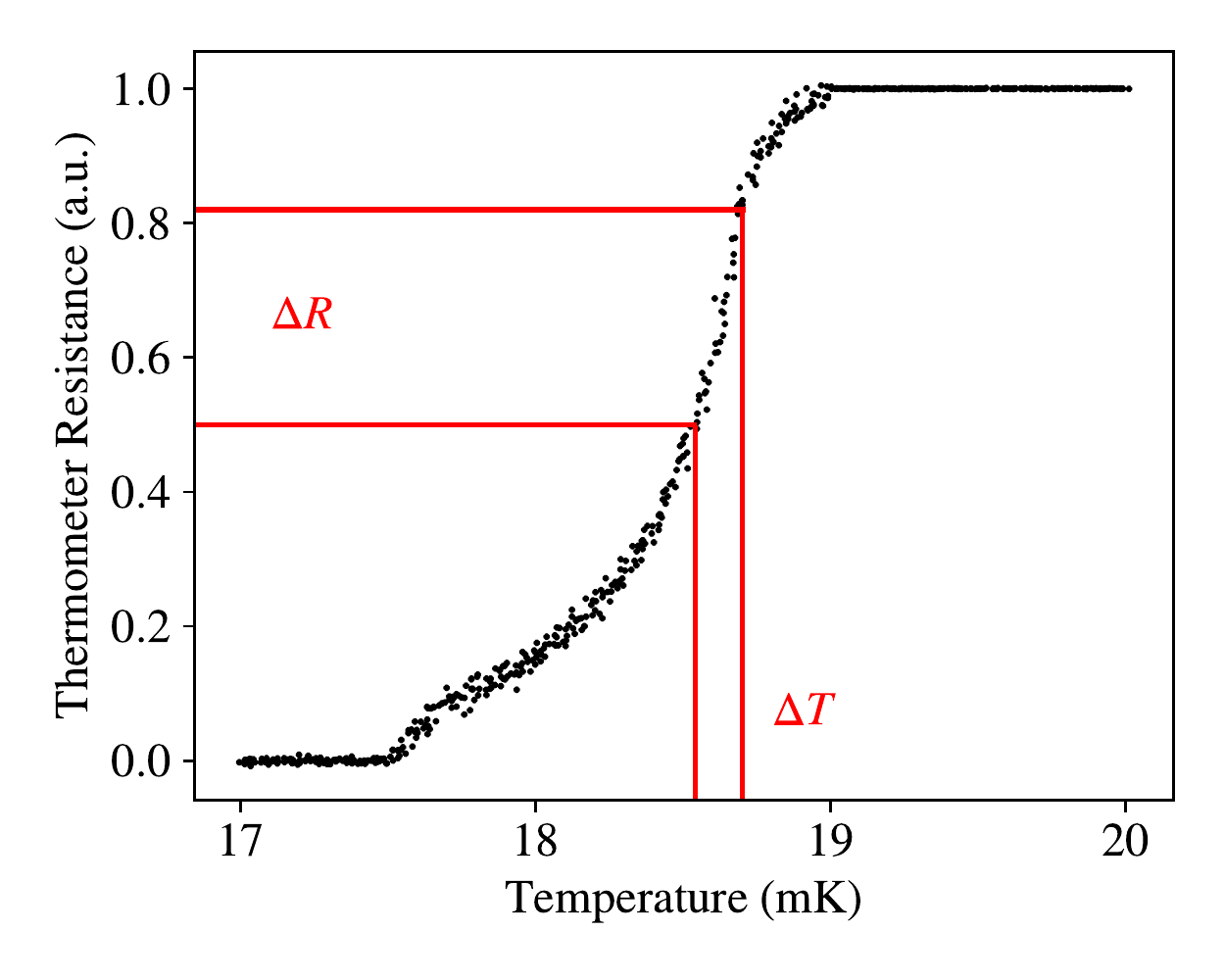}} \quad
\caption{The measured data consists of voltage traces, representing the temperature in the superconducting thermometer. (a) During the measurement process, the pulse-shaped temperature increase in the thermometer (olive) is distorted by additive noise and a non-linear flattening (saturation) of its peak (black). (b) The transition curve of the superconducting thermometer. Operated in between the normal- and superconducting state, a small temperature increase $\Delta T$ causes a measurable increase in resistance $\Delta R$.}
\label{fig:detector}
\end{figure}

Cryogenic detectors are small crystals with a mass of $\mathcal{O}(1- 100)$ g, operated at temperatures below 50 mK. At these low temperatures, an individual particle recoil inside the detector material produces a detectable pulse-shaped heat signal (see Fig. \ref{fig:detector} (a), olive). The crystal is equipped with a sensitive thermometer. In our work we consider a transition edge sensor (TES), i.e. a superconducting film evaporated onto the crystal. The temperature of the superconductor is fine tuned, such that the thermometer is in transition from its normal conducting to its superconducting phase (see Fig. \ref{fig:detector} (b)). This way, the heat signal of the particle recoil causes a measurable increase in the thermometers resistance. For low energy recoils, the transition curve of the thermometer is well approximated with a linear function, i.e. the pulse-shaped heat signature is translated to a pulse-shape voltage signal, only distorted by additive noise. In this regime, the height of the pulse is directly proportional to the energy of the particle recoil. However, for particle recoils with higher energy, the normal-conducting (saturated) region of the thermometer is reached, which we observe as a non-linear flattening of the pulse's peak (see Fig. \ref{fig:detector} (a), black). Throughout a measurement, the shape of the transition curve is continuously measured with injected test signals, making the reconstruction of saturated pulse heights possible. Typical materials used for the detector crystal are $\text{CaWO}_4$, $\text{LiO}_2$, $\text{Al}_2\text{O}_3$,  $\text{Si}$ and $\text{NaI}$ \cite{billard2021direct}. 

\subsection{Used data}

For TES-based detectors, an accurate, parametric model of the response function was derived in Ref. \cite{probst_model_1995}. In Ref. \cite{carrettoni_generation_2010}, a method for the simulation of noise traces with a given noise power spectrum (NPS) is introduced. In Ref. \cite{stahlberg_2020}, the upper end of the detectors saturation curve was approximated with an appropriate fit function. We combine these methods into a simulation of realistic recoil events, which we use exclusively throughout this work. Based on previous work, we expect a good transfer of the results to measured data \cite{Zoller_2016nwp, muehlmann_2019, wagner_2020}, see also Sec. \ref{sec:discussion}. 

We simulate noise traces with a record length of 16384 samples, and a sampling frequency of 25 kHz. The pulse shape is chosen to have a significant signal length of 150 ms and is superposed twice to the noise trace, to augment the desired pile up events (see Fig. \ref{fig:events}). The onset of the first pulse is located in close proximity to the typical trigger position at one fourth of the record window. The second pulse is located on the tail of the first pulse, uniformly distributed between the trigger position and the end of the record window. Both pulse heights are uniformly distributed between 0 and 0.5 V. Finally, the thermometers saturation curve is applied to the voltage trace, flattening the peaks of pulses above 0.25 V. 

The details of the simulation process, including all used parameters, are described in App. \ref{app:sim}. The simulated data set will be made publicly available with the deanonymized version of this work.


\section{Results} \label{sec:results}

\begin{figure}[!t]
\centering
\includegraphics[width=\textwidth]{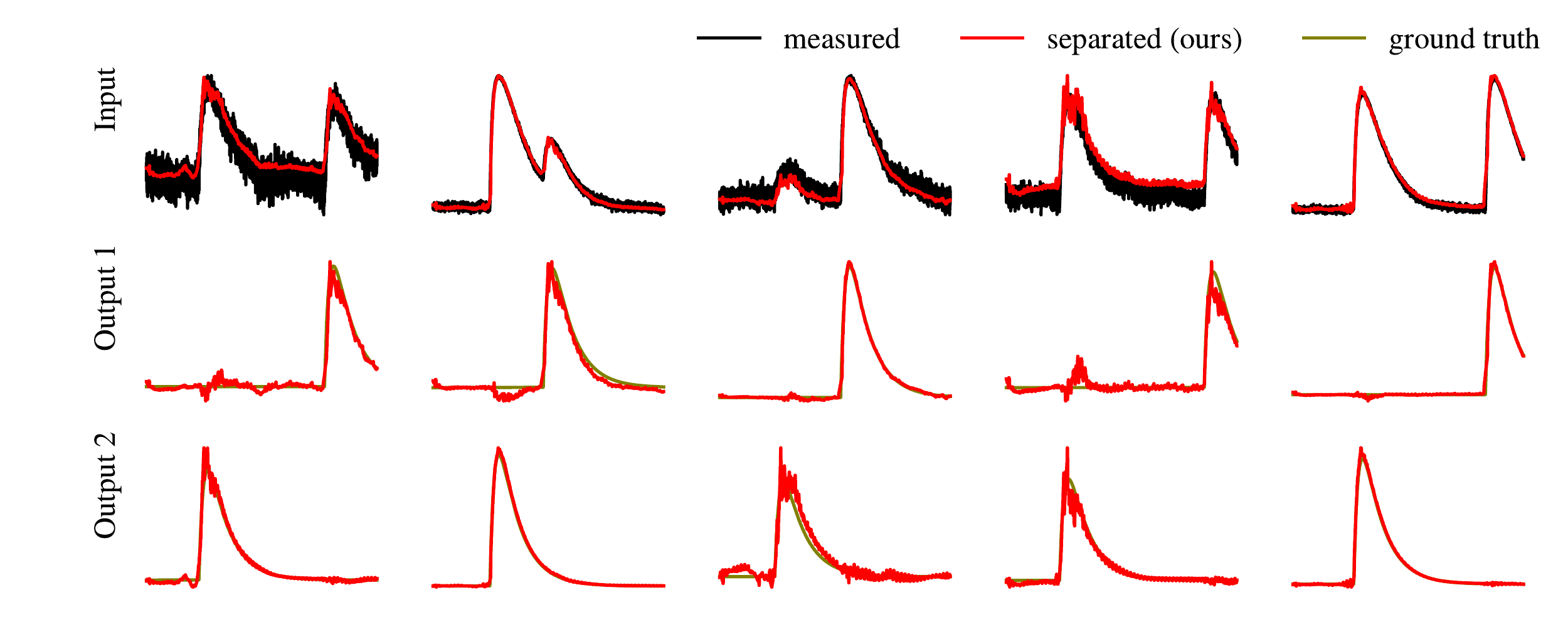}
\caption{The record windows of the recorded particle recoils. The five columns correspond to five independent samples in the data set. The first row shows the time series which are input to the LSTM (black) and the summed outputs after application of the non-linear transition function (red). We observe a good agreement in most samples, with minor distortions for events with low signal-to-noise ratio. The second and third row show the two output traces of the LSTM (red), which are mostly in good agreement with the ground truth, plotted underneath (olive).}
\label{fig:events}
\end{figure}

We use a data set of 60k samples, split into 42k training, 6k validation and 12k test samples. \footnote{For training set sizes > 10k samples, no significant improvement of the results with larger training sets was observed.} The realistic pulse trace (Fig. \ref{fig:events}, black) is the input to our supervised model, the two separate, unsaturated temperature pulses (Fig. \ref{fig:events}, olive) are the training objective. As model, we use a 3-layer LSTM with an input size of 8 values per time step and 200 nodes in each hidden layer. The last layer's hidden states of each time step are linearly mapped to two traces of 8 values each, corresponding to the time step of the two separated temperature pulses.

In a preprocessing step, we perform two transformations on the data. First, all samples are min-max normalized to a value range from 0 to 1. Second, each trace is downsampled by a factor of 32, which leaves to a record length of 512 values. A method, originally described in Ref. \cite{mancuso_method_2019}, that uses the distribution of the maxima of empty noise traces to determine the number of independent samples within the noise of a record window,  leads to approximately the same number (512) of independent samples. This motivates the assumption that with our choice of downsampling, no significant information is lost. With 8 values as simultaneous input to the LSTM, one record window corresponds to 64 time steps.

We train our model with the averaged mean squared error (L2) between outputs and ground truth as loss function. The Gaussian distribution of the additive detector noise speaks in favor of a loss function sensitive to outliers (as the L2), and our decision for this loss function was experimentally confirmed: Others, e.g. the L1 loss function, were explored and brought similar or worse results. The ADAM optimizer with a learning rate of 1e-5 and a batch size of 32 is used, training proceeds for 100 epochs. The best validation score was reached after epoch 65, although already after 15 epochs no further significant improvement on the validation loss score is visible. The final scores were $2.4 \cdot 10^{-4}$ on the test set, and $1.9 \cdot 10^{-4}$ on the training set, which shows no significant overfitting on the training data. The manageable size of the trained model (3.3 MB), together with its fast inference time (Tesla P100 GPU with 12 Gig RAM / Intel(R) Xeon(R) Gold 6138 CPU @ 2.00GHz $\sim$ 0.35 / 5 minutes per training epoch) makes the model suitable for smaller machines. By choosing a reasonable inference batch size of e.g. 1000, this corresponds to the separation of >$3000$ events / second on a single-core CPU.

\begin{figure}[!t]
\centering
\subfloat[]{\includegraphics[width=0.47\textwidth]{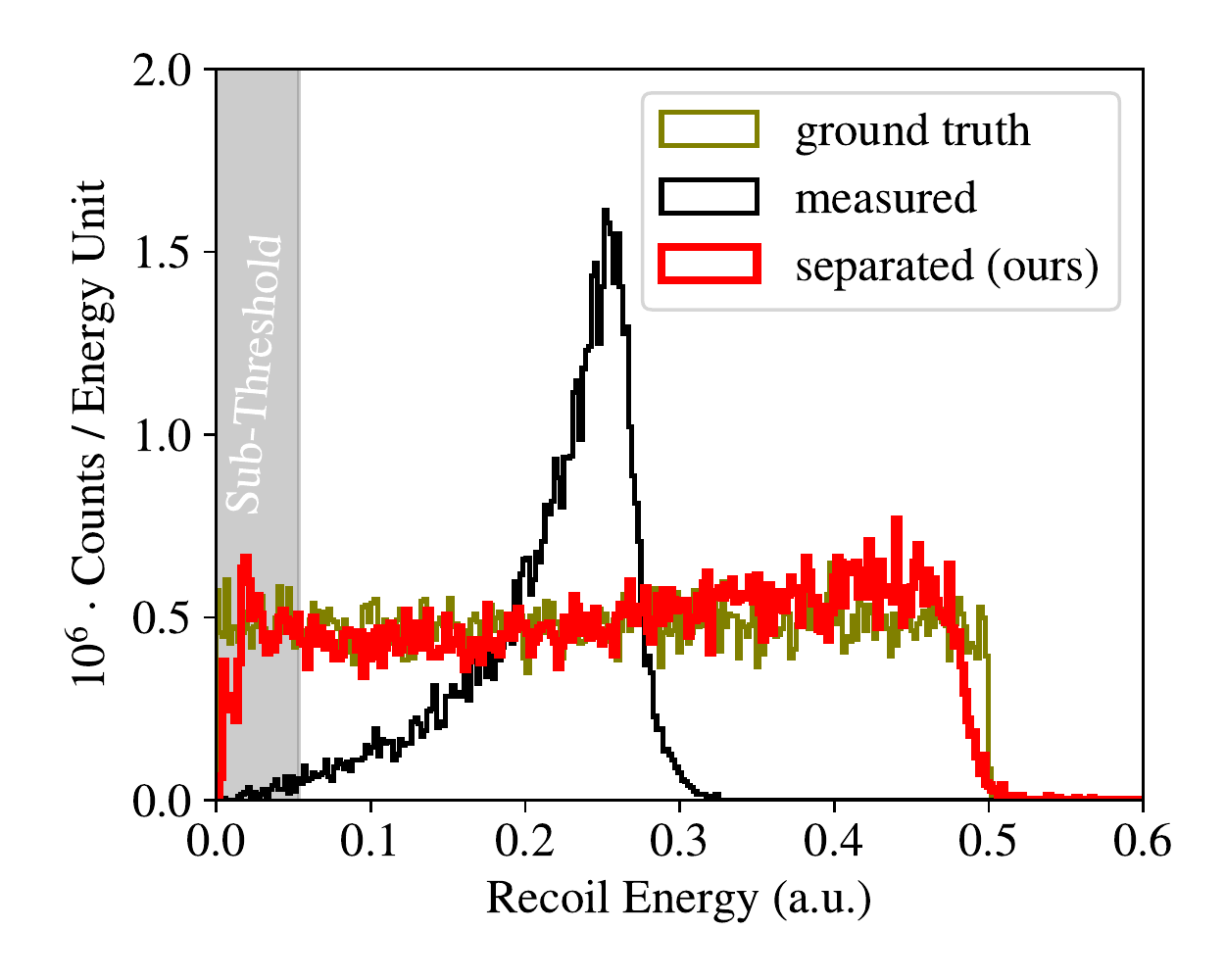}} \quad \quad 
\subfloat[]{\includegraphics[width=0.47\textwidth]{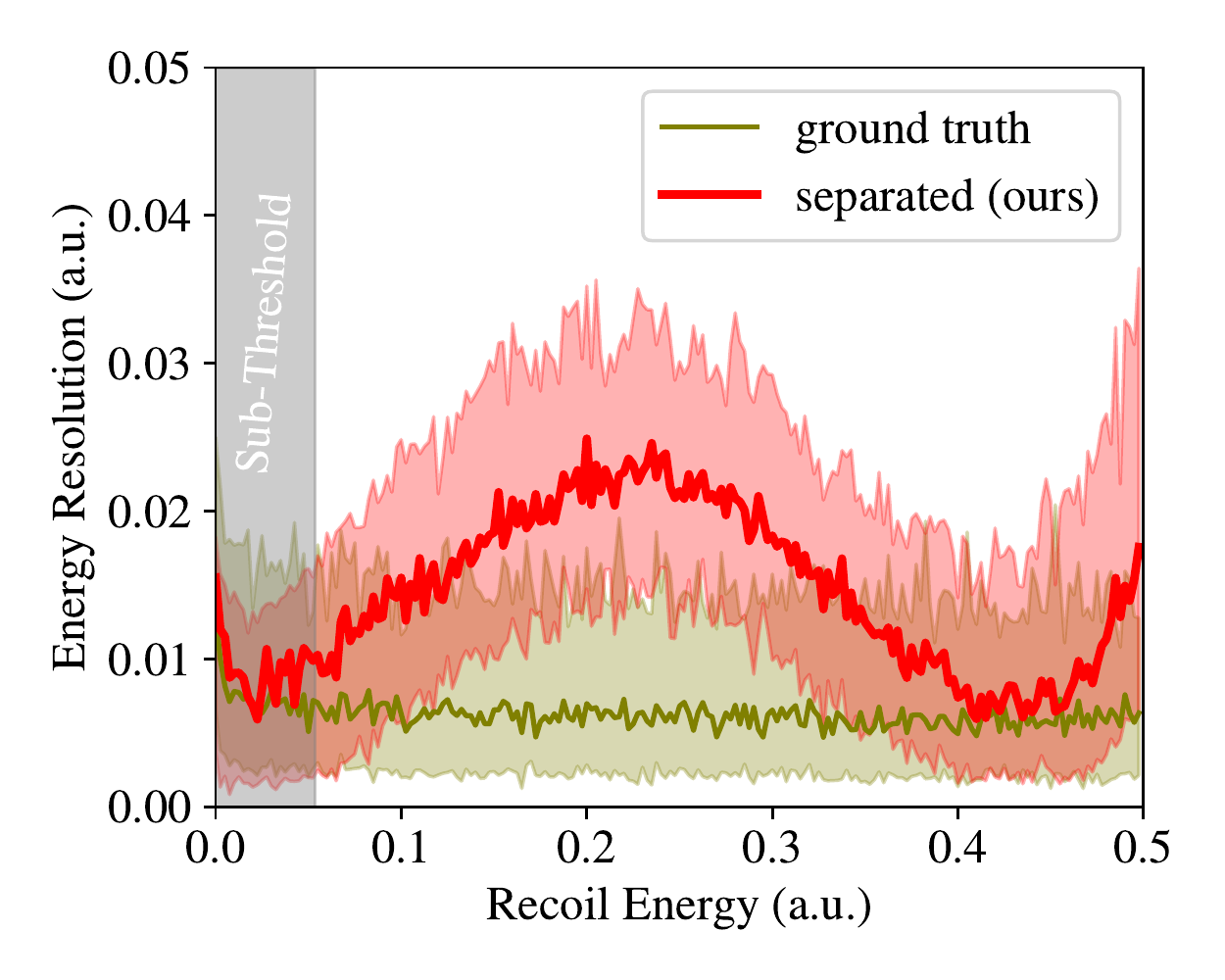}} \quad
\caption{The results of the pulse height reconstruction on an independent test set. (a) The black histogram represents the maximal values of the piled up and saturated traces. These strongly deviate from the uniformly sampled ground truth of the appearing pulse heights (two per record window, olive). We observe a much better agreement of the maximal values of the separated pulse traces (red), with only slight deviations for strongly saturated pulse shapes, close to the upper end of the spectrum. (b) The energy resolution of pulses without pile up and saturation does not depend on the recoil energy, but only on the noise conditions (olive). The energy resolution of the reconstructed pulses does depend on the energy, with local maxima at 0.25 and 0.5, and local minima at 0 and 0.45. This peculiar shape is discussed in Sec. \ref{eres}.}
\label{fig:stats}
\end{figure}

\label{eres} The separation and reconstruction of the individual pulse traces works reasonably well, as shown in Fig. \ref{fig:events}. Minor distortions for pulses with low signal-to-noise ratio are visible, as well as residuals on the separated traces at the position of the first pulse. The impact of the distortions on the energy reconstruction of individual events depends on the choice of the pulse height estimator. While it likely impacts the maximum of the array, this effect might be reduced by commonly used fit- and matched filter methods. In this work, we use the maximum of the array as estimator, to achieve a conservative evaluation of our methods practicability. Therefore we reconstruct as a metric the pulse height spectrum, i.e. the recoil energy spectrum, of the separated pulses from the test set (LSTM predictions, Fig. \ref{fig:stats} (a), red) and compare it with the spectrum of the simulated, individual pulses (ground truth, Fig. \ref{fig:stats} (a), olive) and the piled-up and saturated spectrum (LSTM input, Fig. \ref{fig:stats} (a), black). Additionally, we measure the 0.16, 0.5 and 0.84 quantiles (corresponding to the central value and a one-sigma band in the Gaussian limit) of the deviation of our reconstructed pulse heights from the ground truth value, also called the energy resolution (see Fig. \ref{fig:stats} (b), red). We compare this value with an equivalently calculated value on a similar data set, without pile-up and saturation effects, acting as ground truth (Fig. \ref{fig:stats} (b), olive). 

For both metrics, we reach reasonable agreement of our separated pulse traces with the ground truth: The strongest relative count rate deviation in the separated recoil spectrum from the ground truth (Fig. \ref{fig:stats} (a), red and olive) is $\sim 20\%$ at $0.45$, while it is in the measured spectrum $>300\%$ at 0.25. The strongest relative deviation in the energy resolution from the ground truth (Fig. \ref{fig:stats} (b), red and olive) is a factor 3 at 0.22, corresponding to a relative error in the reconstruction of individual recoil energies of $10 \%$ (LSTM predictions) instead of $3.5 \%$ (ground truth). At threshold, the energy resolution of the LSTM predictions is roughly $25 \%$ higher than that of the ground truth, which is proportional to the necessary increase of the threshold value, to achieve the same noise trigger rate. As expected, the quality of the reconstruction decreases with rising pulse heights, due to the saturation. The observed minimum in the energy resolution at 0.45 (Fig. \ref{fig:stats} (b), red) is a matter of ongoing investigation. One, but likely not the only, contributing effect is of statistical nature and common to all plots of prediction error w.r.t. ground truth: The negative bias of pulses at the upper end of the simulated spectrum (see Fig. \ref{fig:stats} (a), right end of red curve) causes an overdensity of predictions around 0.45. Therefore, events with a true energy around 0.45 feature a lower error, simply because the model predicts these values more often. Overall, the achieved energy resolution is good enough to reconstruct the position of peaks (e.g. for energy calibration) in the spectrum: A Gaussian peak in the ground truth at 0.22 would lead to a peak in the reconstructed spectrum at $0.22 \pm 0.022/\sqrt{N}$, where $N$ is the number of events contributing to the peak. The number of events in a calibration peak is typically high (>> 1000), leading to a small systematic error in the reconstruction of the peaks position. Therefore, our methods makes it possible to extract the information from physics measurements, that otherwise would need to be discarded. We lack the comparison with a state-of-the-art model, as for our application no state-of-the-art has emerged yet.

Finally we want to address the scenario, where the model is applied to a record window with only one pulse, i.e. a regular event. The static architecture of the LSTM model creates, also in this case, two output traces, which are then both interpreted as events. Because the model does not include an automatic discrimination mechanism between regular and pile-up events, one of the output traces creates a noise trigger event that does not correspond to an actual particle recoil. However, in the analysis process usually a lower threshold is imposed on the accepted events, depending on the energy resolution of the detector (Fig. \ref{fig:stats} gray area). By accepting only traces with events above threshold, we can reject all virtual events that appeared in the separation process due to the static number or two output traces. Therefore, the model can be applied to the full data set, without prior discrimination between regular and pile-up events, and without creating additional noise trigger events.

Our trained model and code is public on GitHub \cite{pileupsep}.


\section{Discussion and outlook} \label{sec:discussion}

As shown in Sec.~\ref{eres}, our model performs reasonably well in the task of reconstructing recoil energy spectra from piled-up and saturated measurements, after training on simulated data. For any measurement with a cryogenic detector, the extraction of all parameters necessary for a realistic data simulation is a straight forward procedure. This renders the model useful for practical applications in the current state already. The fast inference time of our model enables real-time inference (processing >3000 event/second vs typical trigger rates $\sim 1$ Hz). However, the choice of a min/max normalization would in this case necessarily be replaced with e.g. a batch normalization layer, to account for the previously unknown min/max values in a non-static data set. There are no publicly available raw data sets of experimentally measured recoil events in cryogenic detectors. Therefore, an evaluation of our model on measured data is still work in progress, in cooperation with the CRESST and COSINUS collaborations. A publication of these results is anticipated. Also, we plan on extending the method for treatment of pile-up with more than two pulse shapes, and pile-up with detector artifacts within the same data set.


\section{Broader impact}

Apart from the previously mentioned experiments CRESST, COSINUS and NUCLEUS, many more rare event searches use cryogenic detectors with pulse-shaped time series data, e.g. the CUORE \cite{PhysRevLett.124.122501}, CUPID \cite{armengaud_cupid-mo_2020}, SuperCDMS \cite{PhysRevLett.127.061801}, EDELWEISS \cite{PhysRevD.99.082003} and RICOCHET \cite{Billard_2017} experiments. Our method can impact the reconstruction of energy spectra from above-ground and calibration runs, potentially decreasing the uncertainty in the energy scales of mentioned experiments. Furthermore, our method can be adapted to separate sensor signals in all application areas of electrical engineering.

\begin{ack}

We thank Wolfgang Waltenberger and Florian Reindl for fruitful discussions and advice on cryogenic experiments and machine learning in particle physics. We thank the Institute of High Energy Physics, the Austrian Academy of Sciences and the Austrian Research Promotion Agency for their support. The computational results presented were obtained using the CLIP cluster. We also thank the reviewers of the Machine Learning and the Physical Sciences workshop at NeurIPS21 for their feedback, which helped significantly in shaping this project.

\end{ack}

\bibliographystyle{unsrtnat}
\bibliography{wagner}

\appendix

\section{Simulation details} \label{app:sim}

\begin{figure}[!t]
\centering
\subfloat[]{\includegraphics[width=0.47\textwidth]{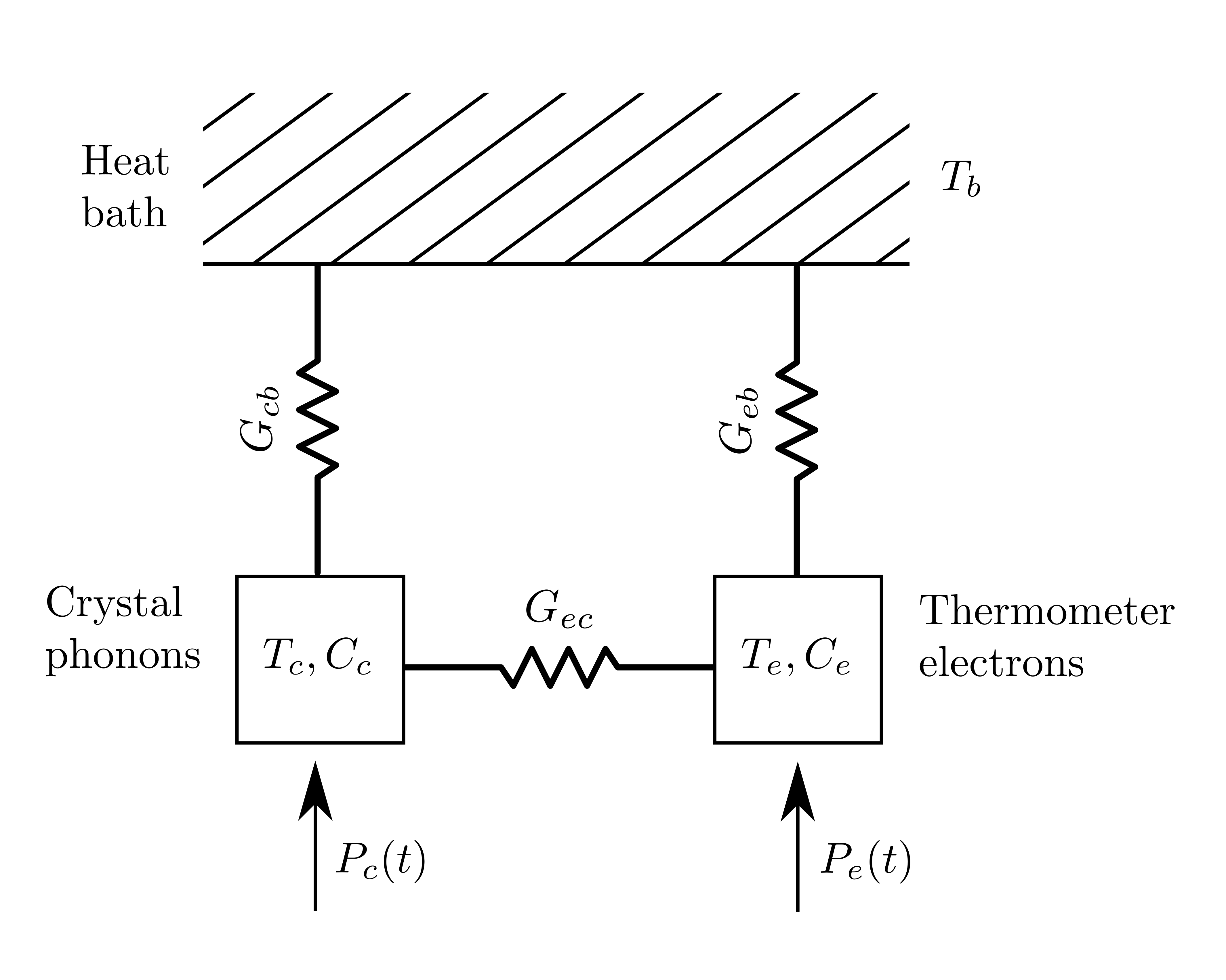}} \quad \quad 
\subfloat[]{\includegraphics[width=0.47\textwidth]{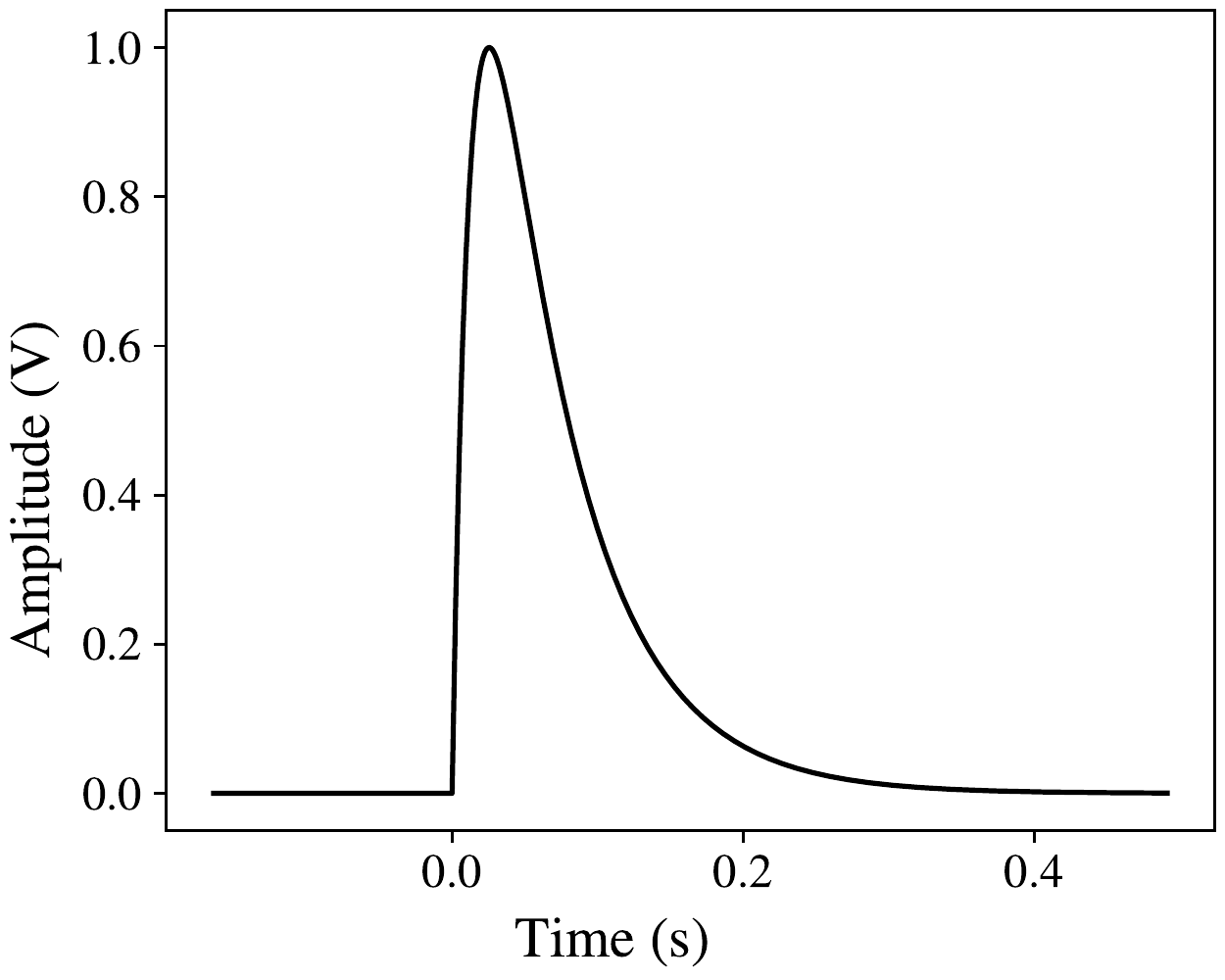}} \quad
\caption{(a) Thermal resistance circuit of the detector: The temperature of the thermometer electrons $T_e$ is the measured quantity, while the temperature of the crystal phonons $T_c$ is a hidden state of the model. The temperature of the heat bath $T_b$ is fixed. The component's heat capacities $C_c$, $C_e$ and the thermal conductivities $G_{cb}$, $G_{eb}$, $G_{ec}$ govern the signal's pulse shape (b).}
\label{fig:thermo}
\end{figure}

In Ref. \cite{probst_model_1995}, the detector is described as a thermal circuit, identifying the crystal phonons and the thermometer electrons as the main components (see Fig. \ref{fig:thermo} (a)), described by their temperature $T_c, T_e$  and heat capacities $C_c, C_e$. Other parts are absorbed in the thermal resistances $G_{cb}, G_{ce}, G_{eb}$, the system is connected to a heat bath regulating the base temperature $T_b$. The temperature of the individual components can be calculated with a system of coupled, ordinary differential equations, depending on the power acting on the crystal and thermometer $P_e, P_c$:

\begin{align}
C_{i} \frac{d T_{i}}{d t}+G_{i b}\left(T_{i}-T_{b}\right)+\sum_{j} G_{j i}\left(T_{j}-T_{i}\right)=P_{i}
\label{eq:system}
\end{align}

The indices $i$ and $j$ correspond to the components $e$ and $c$ respectively. A particle recoil inside the crystal produces a phonon population, a large share of which thermalizes through scattering on the crystals surface, while a small share thermalizes directly inside the thermometer. The waiting time for the thermalization of individual phonons is exponentially distributed, which leads to exponentially decaying power inputs in both the crystal and the thermometer. The subsequent heat transport from the crystal to the thermometer causes a second pulse component, introducing three time constants of the detectors response signal in total: $\tau_{in}$, the time until thermalization inside the thermometer, $\tau_{n}$ the time needed for the heat transport from the crystal to the thermometer and $\tau_{t}$, the overall relaxation time of the system. Solving Eq. \ref{eq:system} with an Ansatz of an exponential decay for the power inputs leads to a solution for the heat increase in the thermometer, which describes the detectors response signal very well:

\begin{align}
\Delta T_{e}(t)=\Theta(t)\left[A_{n}\left(\mathrm{e}^{-t / \tau_{n}}-\mathrm{e}^{-t / \tau_{i n}}\right)+A_{t}\left(\mathrm{e}^{-t / \tau_{t}}-\mathrm{e}^{-t / \tau_{n}}\right)\right]
\label{eq:pulsemodel}
\end{align}

The amplitudes of the individual signal components $A_n, A_t$, as well as the time constants, depend on the thermal resistances and heat capacities. An exemplary pulse shape is shown in Fig. \ref{fig:thermo} (b). For a measurement done with cryogenic detectors, these parameters can be identified by a fit of Eq. \ref{eq:pulsemodel} to a standard event, i.e. the superposition of several typical recoil events. Most recoil events follow this characteristic pulse shape, only for high energy events a nonlinear saturation effect occurs. Extensions of the pulse shape model, for detectors with more than two thermal components, are derived in Ref. \cite{Zema:2020mkm}.

\begin{figure}[!t]
\centering
\subfloat[]{\includegraphics[width=0.47\textwidth]{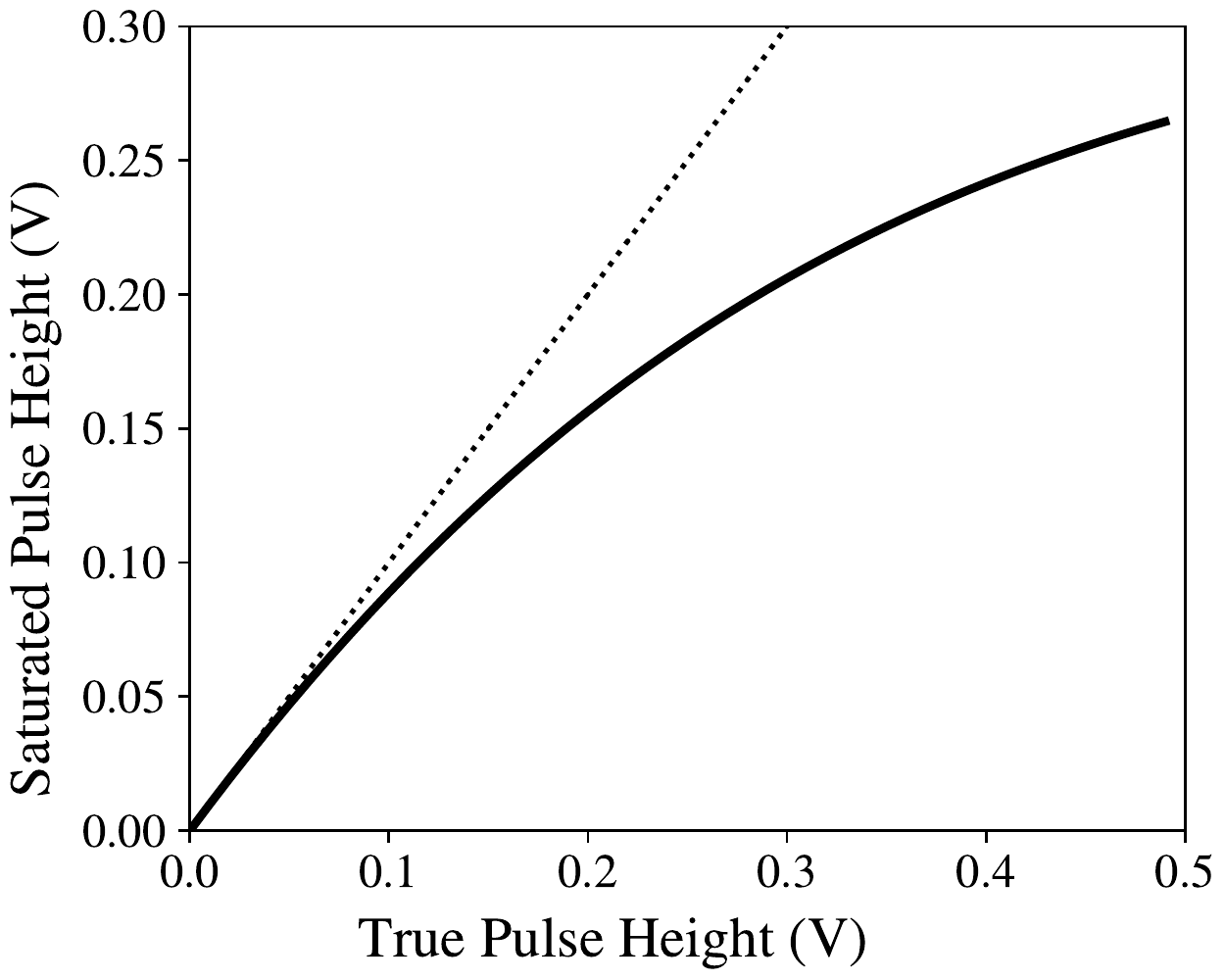}} \quad \quad 
\subfloat[]{\includegraphics[width=0.47\textwidth]{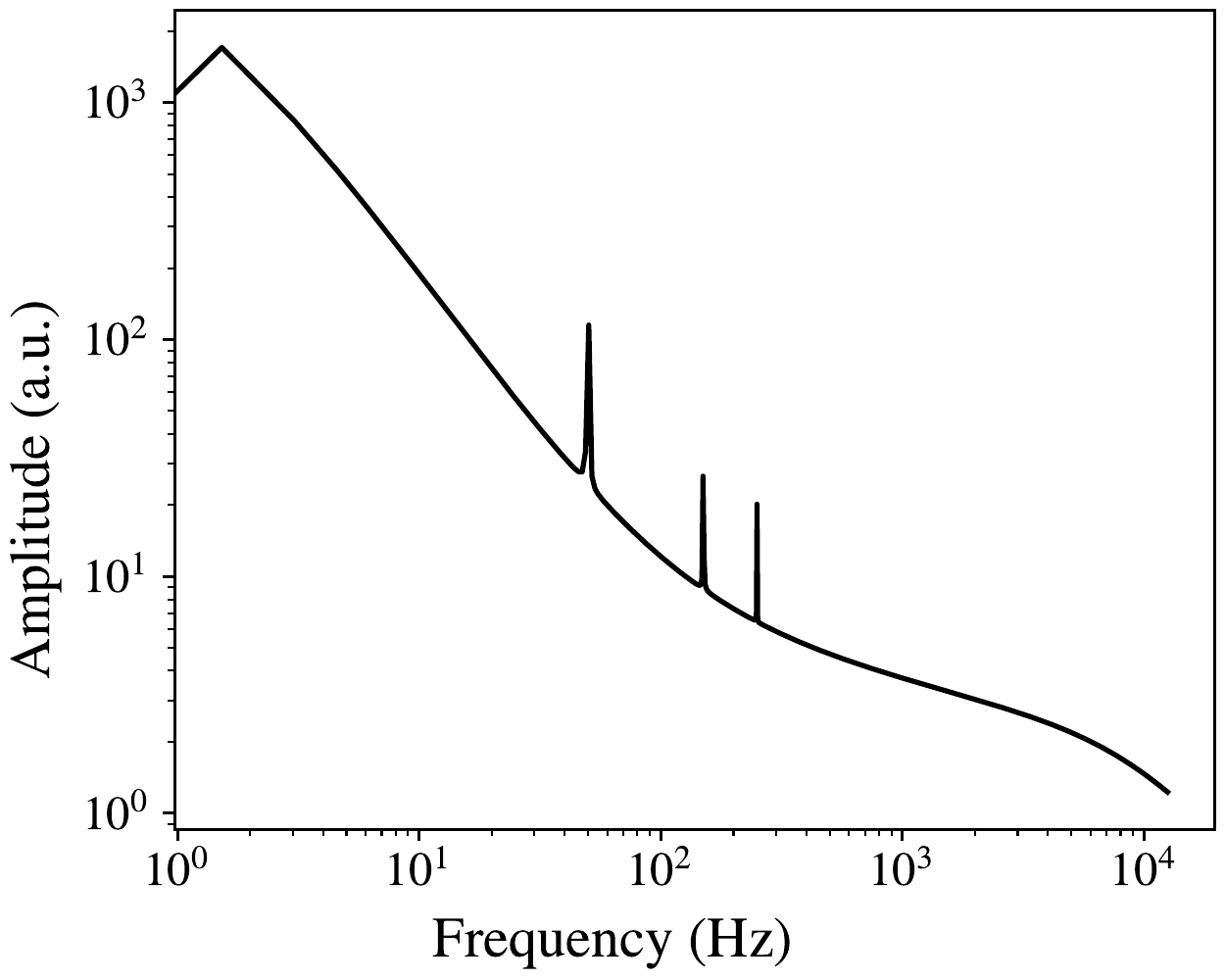}} \quad
\caption{(a) The detector's saturation causes a flattening of the peaks of signal pulses. This saturation can be described with a generalized logistics function (Eq. \ref{eq:log}). The variable $Y$ corresponds to the saturated pulse height, $t$ to the true pulse height. (b) A simulated noise power spectrum, built from a superposition of two $1/f^{\alpha}$ functions, with different parameters $\alpha$ that describe the pink and white noise components. Disturbances from main voltage at 50, 100 and 150 Hz are added, the constant component at 0 Hz is set to zero. Frequencies above 10 kHz are cut with an anti-aliasing low pass filter.}
\label{fig:nps}
\end{figure}

The saturation depends on the superconductor's transition curve and was in Ref. \cite{stahlberg_2020} modeled with a generalised logistic function (see Fig. \ref{fig:nps} (a)):

\begin{align}
Y(t)=A+\frac{K-A}{\left(C+Q e^{-B t}\right)^{1 / \nu}},
\label{eq:log}
\end{align}

where $K, C, Q, B$ and $\nu$ are free fit parameters and $A$ is derived from the condition $Y(0) = 0$. The variable $t$ describes the true, unsaturated height of a signal pulse, $Y$ describes the saturated pulse height. 

For the simulated data we used throughout this work, we superpose simulated noise baselines and the pulse shape model, and apply the generalized logistic function point-wise.

The simulation of noise baselines follows the method proposed in Ref. \cite{carrettoni_generation_2010}, with an additionally added polynomial drift structure that often appears in measured noise traces. A noise power spectrum is built from a parametric model, reflecting the typical $1/f^\alpha$ shape, the interference from mains voltage and an anti-aliasing low pass filter (see Fig. \ref{fig:nps} (b)).

For the pulse shape model we used the parameters:

$$ \tau_t = 0.05815608, \tau_{in} = 0.02059209, \tau_n = 0.01395427, A_n = 0.35, A_t = 1.97. $$

For the generalised logistic function we used the parameters:

$$ K = 10, C = 3, Q = 0.5, B = 5, \nu = 0.5 $$

\end{document}